\documentclass[aps,prl,preprint,groupedaddress]{revtex4-1}

\usepackage{graphicx}
\begin{document}

\begin{flushright}
OSU-HEP-16-03\\
UMD-PP-016-004
\end{flushright}


\title{\large Limiting Equivalence Principle Violation and Long--Range\\[-0.08in] Baryonic Force from Neutron-Antineutron Oscillation}]


\author{\bf  K.S. Babu$^a$ and Rabindra N. Mohapatra$^b$}
\affiliation{$^a$Department of Physics, Oklahoma State University, Stillwater, OK, 74078, USA}
\affiliation{$^b$Maryland Center for Fundamental Physics and Department of Physics, University of Maryland, College Park, Maryland, 20742, USA}


\date{\today}

\begin{abstract} We point out that if the baryon number violating neutron-antineutron oscillation is discovered, it would impose strong limits on the departure from Einstein's equivalence principle at a level of one part in $10^{19}$. If this departure owes its origin to the existence of long-range forces coupled to baryon number $B$  (or $B-L$), it would imply very stringent constraints on the strength of gauge bosons coupling to baryon number current. For instance, if the force mediating baryon number has strength $\alpha_B$ and its range is larger than a megaparsec, we find the limit to be $\alpha_B \leq 2\times 10^{-57}$, which is much stronger than all other existing bounds. For smaller range for the force, we get slightly weaker, but still stringent bounds by considering the potential of the Earth and the Sun.
\end{abstract}


\maketitle

\section{1. Introduction}
Equivalence principle is one of the pillars of Einstein's general relativity. The success of general relativity has therefore led, over the years, to  many attempts to search for deviation from this principle. These attempts have so far been unsuccessful and have provided very stringent upper limits on any possible deviation. One way to interpret a deviation from equivalence principle is to assume that there exist long-range forces with sub-gravitational strengths and the above mentioned upper limits are then reflections on the strength of these new long-range forces. A very well known early example of such an interpretation is the work of Lee and Yang~\cite{yang} who obtained a limit  $\alpha_B \leq 6\times 10^{-44}$ on the strength of the long-range force coupled to baryon number.

In this brief note, we point out that if the baryon number violating process of neutron to antineutron oscillation~\cite{nnbar} is observed, regardless of the level at which it is discovered, it will put an upper limit on the deviation of equivalence principle for neutrons and antineutrons. If this deviation is attributed to the existence of a $U(1)_B$ (or $U(1)_{B-L}$) local symmetry coupled to baryon number with an associated long-range force, we find very stringent limits on the strength of this 
long-range force (denoted by $\alpha_B$). The limits depend on the range of the force. The most stringent limit arises in the case when the the range of the force is larger than 100 megaparsec (Mpc), and is found to be $\alpha_B\leq 10^{-54}$, which is significantly stronger than that derived by Lee and Yang.

\section{2. Neutron- antineutron oscillation and bound on departure from equivalence principle}
The basic equation that we use in our discussion is the quantum mechanical evolution of two state system for $n$ and $\bar{n}$ in the presence of an external field that distinguishes between neutrons and antineutrons:
\begin{eqnarray}
\frac{d}{dt}\left(\begin{array}{c} n \\ \bar{n}\end{array}\right)~=~\left(\begin{array}{cc} M_1 & \delta \\ \delta & M_2\end{array}\right)\left(\begin{array}{c} n \\ \bar{n}\end{array}\right).
\end{eqnarray}
If we start with an initial beam of neutrons, the probability that an antineutron beam will appear after a transit time of $t$ is given by:
\begin{eqnarray}
P_{n-\bar{n}}~=~\frac{\delta^2}{\Delta M^2+\delta^2}{\rm sin}^2 \frac{\sqrt{\Delta M^2+\delta^2} ~t}{\hbar}
\end{eqnarray}
where $\Delta M=M_2-M_1$. This difference could arise from a magnetic field ~\cite{MM2} or from nuclear forces, for example. In our discussion here, it will owe its origin to departure from equivalence principle and/or new long range forces that distinguish between neutrons and antineutrons. For a transit time $t$, the condition for observability of $n-\bar{n}$ oscillation~\cite{MM2} is that $\Delta M t \leq 3\times 10^{-24}$ GeV-sec. For transition time of order of one second, which is what realistic experimental setups can achieve with current technology, this condition would imply $\Delta M \leq 3\times 10^{-24}$ GeV as a generous upper limit.  Thus, the observation of $n-\bar{n}$ oscillation
will impose a constraint on the strength of the forces that are responsible for causing the mass difference. This constraint was used recently to obtain  a limit on possible violation of Lorentz invariance~\cite{BM}.

To obtain the limit on the departure from equivalence principle for neutrons and antineutrons, all we have to do is to calculate $\Delta M$. We adopt the following parametrization for this purpose. Let us consider a source of gravitational potential of mass $M$ which is at a distance $r$ from the neutrons in the experiment searching for $n-\bar{n}$ oscillation. Assuming that the force causing the departure to be long-range,  we can parameterize the departure from equivalence principle for neutrons given by the potential  $\alpha_n \frac{GMm}{r}e^{-r/R_0}$ and antineutrons by $\alpha_{\bar{n}} \frac{GMm}{r}e^{-r/R_0}$ (where $m$ is the mass of the neutron).  Then we obtain
\begin{eqnarray}
\Delta M~=~(\alpha_n-\alpha_{\bar{n}})\frac{GMm}{r}e^{-r/R_0}.
\end{eqnarray}
Consideration of different astrophysical sources, which will have different $M$ and different $r$, we can get different limits on $(\alpha_n-\alpha_{\bar{n}})$. Below we summarize the different limits by considering the Earth, Sun and the superclusters. Clearly, the validity of the limits will depend on the range of the forces.

\subsection{2.a~~Superclusters limit} We consider a typical supercluster such as Virgo which is at a distance of 16.5 Mpc and has a mass of $2.4\times 10^{45}$ kg. For this we get $\frac{GMm}{r}\simeq 3.6\times 10^{-6}$ GeV. Using the fact that the corresponding $\Delta M \leq 3\times10^{-24}$ GeV (required if $n-\bar{n}$ oscillation is observed), we get the bound
\begin{eqnarray}
(\alpha_n-\alpha_{\bar{n}})\leq 10^{-18}.
\end{eqnarray}
This limit on equivalence principle violation is more stringent than any known at the moment for baryons~\cite{equiv}. The results of Dicke et. al. and Braginsky et al. are at the level of $10^{-12}$~\cite{brag}.  The most stringent limit from $K^0-\bar{K^0}$ oscillations seems to be comparable to ours~\cite{kenyon},
$(\alpha_K-\alpha_{\bar{K}})\leq 2.6\times 10^{-18}$.

\subsection{2.b ~Limit from Earth's gravitational field} If the range $R_0$ of the equivalence principle violating effect is $\sim 10,000$ km, then the supercluster limits will not apply (due to the  $e^{-\frac{r}{R_0}}$ suppression factor for $r \gg R_0$), but there should be a limit by considering the effect of the Earth. Using the mass of the Earth as $6\times 10^{24}$ kg and the radius of the Earth as $R_E=6384$ km, we estimate that the Earth's effect leads to $(\alpha_n-\alpha_{\bar{n}})\leq 4\times 10^{-15}$.

\section{3.~Gauged Baryon number and limit on long-range baryonic force from observation of $n-\bar{n}$ oscillation}
Gauging baryon number has been considered for a long time as way to understand the conservation of baryon number in the universe~\cite{yang}. In particular, Lee and Yang derived a limit on the strength of the effective baryon number force $\alpha_B$ to be at the level of $10^{-47}$ if we parameterize the resulting  potential as
\begin{eqnarray}
V_B(r)~=~\alpha_B \frac{N_AN_B}{r} e^{-r/R_0}
\end{eqnarray}
where $N_{A,B}$ are the baryon numbers of the two objects between which the above potential is effective and $R_0$ is the range of the force. Understanding baryon asymmetry of the universe seems to require as one of its ingredients that baryon number be violated. This has led to a new class of models where local baryon number symmetry is spontaneously broken~\cite{recent}. Similar situation also happens for $B-L$ violation~\cite{MM}. Typically, in these models, one assumes that the corresponding gauge coupling $g_B$ is is of order $\sim 0.1-1$- so that for spontaneously generated vacuum expectation value $v_B\sim TeV$, the resulting force is short-range and is not relevant in the discussion of violation of equivalence principle at macroscopic distances. In this section, we will adopt a somewhat different point of view where  even though the local baryon number symmetry is broken spontaneously at a few hundred GeV to TeV scale, the associated gauge coupling is very small. For example, if the gauge coupling is $\leq 10^{-25}$, the range of the force with $v_B=1$ TeV is larger than the Earth radius and will in principle affect the equivalence principle between neutron and antineutron.

Note that since in our theory, neutron-antineutron oscillation is allowed to occur at an observable rate, we must have Feynman diagrams for $\Delta B=2$ processes, which give strengths at the quark level of $10^{-28}$ GeV$^{-5}$.  In beyond the standard model scenarios, $n-\bar{n}$ oscillation arises from the six quark operator $(udd)^2$ and its strength in a typical $B-L$ violating theory~\cite{MM} is given by $G_{\Delta B = 2}\sim \frac{\lambda f^3 v_{BL}}{M^6_\Delta}$. Thus we can have observable $n-\bar{n}$ oscillation by choosing the corresponding Yukawa couplings $f$ and Higgs masses $M_{\Delta}$ appropriately for TeV-scale $v_B$. Important to note that in the theory of the type described in ~\cite{MM}, the $\Delta B=2$ diagram does not involve gauge couplings. Thus we can take the theory of Ref.~\cite{MM},  and make the gauge coupling extremely tiny so that it produces corrections to equivalence principle and then check what would be an upper bound on the gauge coupling in this domain of parameters.

Following the procedure above, we find that the neutron and antineutron experience equal and opposite long-range forces from an astrophysical object. Considering the effect of the Earth, we find that the equivalence principle violating parameter $(\alpha_n-\alpha_{\bar{n}})$ is given by
\vspace*{-0.02in}
\begin{eqnarray}
\alpha_n-\alpha_{\bar{n}}~=~\frac{2\alpha_B N^{Earth}_B}{m_nR_{Earth}}\sim 1.2\times 10^{+29}\alpha_B.
\end{eqnarray}
\vspace*{-0.02in}
\noindent Requiring that $n-\bar{n}$ oscillation be observable in the presence of this effect implies that $\alpha_n-\alpha_{\bar{n}}\leq 3\times 10^{-24}$ leading to $\alpha_B \leq 2.5 \times 10^{-53}$, which means that the corresponding gauge coupling $g_B \equiv \sqrt{4\pi \alpha_B}\leq 1.7\times 10^{-26}$. This implies a range $R_0\geq 10^{9}$ cm which exceeds the Earth radius.  This is already a much stronger bound than any known to date~\cite{heeck}.

This bound becomes even stronger, if we apply the same considerations to the Sun. First note that this would require that the gauge coupling be less than $10^{-30}$. Using the mass of the Sun which is $2\times10^{30}$ kg and Earth-Sun distance $\sim 1.5\times 10^{13}$ cm., we get $\alpha_B \leq 10^{-54}$ and hence $g_B\leq 3\times 10^{-27}$. For consistency with range requirement, we must take the symmetry breaking scale $v_B\sim 10$ GeV. 

Coming to the case of Virgo supercluster, where mass and distance are already mentioned, applying similar arguments (if the range of the force $R_0$  is larger than  $10^{26}$ cm), we obtain $\alpha_B \leq 2\times 10^{-57}$ leading to $g_B\leq 1.2\times 10^{-28}$. Clearly to get this kind of range, we must have the symmetry breaking scale to be less than few eV. Such small vev, to be consistent with current limits on the strengths of $n-\bar{n}$ oscillation will require making some parameters in the model small.  Our goal here is not to explore the naturalness of the theory but rather to pursue the phenomenological implications. 

\begin{figure}[h!]
	\centering
\includegraphics[scale=0.7]{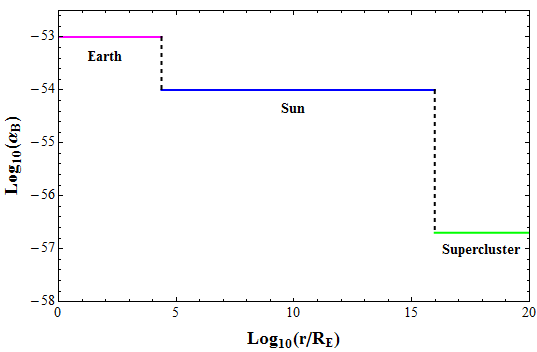}
\caption{Limits on $\alpha_B$ that would result from observation of $n-\bar{n}$ oscillation in the presence of a long-range
baryonic force.  Here $R_E = 6.384 \times 10^8$ cm is the radius of the Earth, and $r$ is the distance between the experiment and the astrophysical source.}
\label{fig1}
\end{figure}

We have summarized in Fig. \ref{fig1} the constraints from long-range baryonic forces that arise from the Earth, the Sun and
superclusters on the strength of the $B$ or $B-L$ gauge interaction $\alpha_B$. 

We point out that if instead of gauged baryon number, we consider a force coupled to gauged $B-L$, we will get a slightly weaker bound since typical astrophysical objects will contain hydrogen and helium atoms in comparable numbers; however the hydrogen atom has zero $B-L$ whereas the Helium atom has $B-L=2$. The factor weakening the bound will depend on the relative content of these two atoms in the astrophysical object.

\section{4. Summary} In summary, in this brief note we have pointed out that observation of neutron-antineutron oscillation, in addition to providing a key window into physics beyond the standard model and possibly solving the baryon asymmetry problem, can also provide insight into violation of equivalence principle as well as limits on the strength of long-range baryonic  gauge forces.  It is important to point out that to obtain the limits discussed above, one has to carry out the search for and observe free neutron oscillation and not a $\Delta B=2$ transition in a nucleus, where such tiny effects are masked by the larger nuclear potential difference affecting the neutron and the antineutron.  It may also be worth noting that, if neutron oscillation inside a nucleus is discovered and no $n-\bar{n}$ oscillation at the same level is found in free neutron oscillation search, that could be evidence for the existence of violation of equivalence principle and/or existence of baryonic long range forces. These results should provide additional impetus to carry out the search for free neutron oscillation in the laboratory.

\section*{Acknowledgement} The work of KSB is supported by the US Department of Energy Grant No. de-sc0016013 and the
work of RNM is supported by the US National Science Foundation Grant No. PHY-1315155. KSB wishes to thank CETUP* 2016 workshop in Lead, SD, where part of this work was done for hospitality and support.

\vspace{.1in}

\noindent{\it Note added:} After this work was completed and presented at the CPT16 meeting in Bloomington, Indiana (June 20-24, 2016), it was brought to our attention by W.M. Snow that a similar work is in progress by Z. Berezhiani and Y. Kamyshkov (to appear).

\end{document}